\documentclass[smus]{snow}
\usepackage{graphics,epsfig,rotating}

\begin{document}

\title{Leptoquark Production and Identification in High Energy 
$\mu^+\mu^-$, $e^+ e^-$ and $e\gamma$ Collisions\thanks{This research 
was supported in part by the Natural Sciences and Engineering 
Research Council of Canada.}}

\author{Michael A. Doncheski$^1$ and Stephen Godfrey$^2$ \\ 
{\it $^1$Department of Physics, Pennsylvania State University, Mont Alto, PA 
17237 USA } \\
{\it $^2$Ottawa-Carleton Institute for Physics} \\
{\it Department of Physics, Carleton University, Ottawa CANADA, K1S 5B6} }

\maketitle

\thispagestyle{empty}\pagestyle{empty}

\begin{abstract} 
We study single leptoquark production in $e^+e^-$, $e\gamma$, and 
$\mu^+\mu^-$ collisions due to the quark content of the photon.  
Leptoquarks can be produced in substantial numbers for masses very 
close to the centre of mass energy of the collider which results in 
equivalently high discovery limits.  
Using polarization asymmetries in an $e\gamma$ collider
the ten different types of 
leptoquarks listed by Buchm\"uller, R\"uckl and Wyler
can be distinquished from one another for leptoquark 
masses essentially up to the kinematic limit.
Thus,  if a leptoquark 
were discovered an $e\gamma$ collider could play a crucial role in 
determining its origins.

\end{abstract}

\section{Introduction}
There is considerable interest in the study of leptoquarks (LQs) --- colour
(anti-)triplet, spin 0 or 1 particles which carry both baryon and lepton 
quantum numbers.  Such objects appear in a large number of extensions 
of the standard model such as grand unified theories, technicolour, 
and composite models.  
Quite generally, the signature for leptoquarks is very
striking: a high $p_{_T}$ lepton balanced by a jet (or missing $p_{_T}$ 
balanced by a jet, for the $\nu q$ decay mode, if applicable).  
Searches for leptoquarks have been performed by the H1 \cite{H1} and 
ZEUS \cite{ZEUS}
collaborations at the HERA $ep$ collider, by the D0 \cite{D0} 
and CDF \cite{CDF} 
collaborations at the Tevatron $p\bar{p}$ collider, and by the ALEPH 
\cite{ALEPH}, 
DELPHI \cite{DELPHI}, L3 \cite{L3}, and OPAL\cite{OPAL} 
collaborations at the LEP $e^+e^-$ collider.  Discovery limits have 
also been obtained for the LHC\cite{LHC}. 
In this report  we consider single 
leptoquark production in $\mu^+\mu^-$, $e^+e^-$, and $e\gamma$
collisions which utilizes the quark 
content of either a backscattered laser photon for the $e\gamma$ case 
or a Weizacker-Williams photon radiating off of one of the 
initial leptons for the $\mu^+ \mu^-$ or $e^+e^-$ cases
\cite{DG1,DG2,eboli,BLN,HP,FC,misc}.  
This process offers the advantage of a much higher 
kinematic limit than the LQ pair production process, is independent of 
the chirality of the LQ, and gives similar results for both scalar and 
vector leptoquarks.  
In the first part of this report 
we use single leptoquark production to estimate the discovery limits at 
future high energy $\mu^+\mu^-$, $e^+e^-$, and $e\gamma$ colliders.  

Although the discovery of a leptoquark would be dramatic evidence 
for physics beyond the standard model it would lead to the question 
of which model the leptoquark originated from.
Given the large number of leptoquark types 
it would be imperative to measure its properties to answer this 
question.  
There are 10 distinct leptoquark types which have been classified by
Buchm\"uller, R\"uckl and Wyler (BRW) \cite{buch}:
$S_1$, $\tilde{S}_1$ (scalar, iso-singlet); 
$R_2$, $\tilde{R}_2$ (scalar, iso-doublet); $S_3$ (scalar, iso-triplet); 
$U_1$, $\tilde{U}_1$ (vector, iso-singlet); $V_2$, $\tilde{V}_2$ (vector, 
iso-doublet); $U_3$ (vector, iso-triplet).  
The production and corresponding 
decay signatures are quite similar, though not identical, and have been 
studied separately by many authors. The question arises as to how to 
differentiate between the different types.  In the second part of this 
report we show how a polarized 
$e\gamma$ collider can be used 
to differentiate the LQs. (ie. a polarized $e$ beam, like SLC,
in conjunction with a polarized-laser backscattered photon beam.)
This process takes advantage of the hadronic component of the photon 
which is important and cannot be neglected \cite{photont}.

\section{Leptoquark Production}

The process we are considering is shown if Fig. 1.  The parton level 
cross section for scalar leptoquark production is trivial, given by: 
\begin{equation}
\sigma(\hat{s})=\frac{\pi^2 \kappa \alpha_em}{M_s} 
                \delta(M_s - \sqrt{\hat{s}})
\end{equation}
where we have followed the convention adopted in the 
literature\cite{BLN} where the leptoquark couplings are replaced 
by a generic Yukawa coupling $g$ which is scaled to electromagnetic 
strength $g^2/4\pi=\kappa \alpha_{em}$.
The cross section for vector leptoquark production is a factor of two 
larger.
We only consider generation diagonal leptoquark couplings so that
only leptoquarks which couple to electrons can be produced in 
$e\gamma$ (or $e^+e^-$) collisions while for the $\mu^+\mu^-$ collider only 
leptoquarks which couple to muons can be produced.
Convoluting the parton level cross section with the quark 
distribution in the photon one obtains the expression
\begin{eqnarray}
\sigma(s) & = & \int f_{q/\gamma}(z,M_s^2) \hat{\sigma}(\hat{s}) dz 
                \nonumber \\
& = & f_{q/\gamma}(M_s^2/s,M_s^2) 
      \frac{\mbox{$2\pi^2\kappa \alpha_{em}$}}{\mbox{$s$}}.
\end{eqnarray}
The cross section depends on the LQ charge since 
the photon has a larger $u$ quark content than $d$ quark content.
We note that the interaction Lagrangian used 
in Ref. \cite{HP} associates a factor $1/\sqrt{2}$ with the 
leptoquark-lepton-quark coupling.  Thus, one should compare our 
results with $\kappa$ to those in Ref. \cite{HP} with $2\kappa$.
We give results with $\kappa$ chosen to be 1.  

\begin{figure}[b]
\leavevmode
\centerline{\epsfig{file=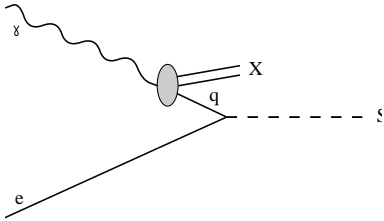,width=5.0cm,clip=}}
\caption{The resolved photon contribution for leptoquark production in 
$e\gamma$ collisions.}
\end{figure}

For $e\gamma$, $e^+e^-$, and $\mu^+\mu^-$ colliders
the cross section is obtained by 
convoluting the expression for the resolved photon contribution to 
$e \gamma$ production of leptoquarks, Eqn. (2), with, as appropriate,
the backscattered laser photon distribution \cite{bsl1} or the 
Weizs\"acker-Williams effective photon distribution:
\footnote{The effective photon distribution from muons is 
obtained by replacing $m_e$ with $m_\mu$.}
\begin{eqnarray}
\sigma(\ell^+ \ell^- \rightarrow X S) & = & \frac{2 \pi^2 \alpha_{em}\kappa}{s} 
    \int_{M_s^2/s}^1 \frac{dx}{x} f_{\gamma/\ell}(x,\sqrt{s}/2) \nonumber \\
  & & 	 \qquad \times    f_{q/\gamma}(M_s^2/(x s), 
M_s^2).
\end{eqnarray}

Before proceeding to our results we consider possible backgrounds 
\cite{DGP}.  The leptoquark signal consists of 
a jet and electron with balanced transverse momentum and 
possibly activity from the hadronic remnant of the photon.  
The only serious background is a hard scattering of a quark inside the 
photon by the incident lepton via t-channel photon exchange; $eq \to 
eq$.  By comparing the invariant mass distribution for this background 
to the LQ cross sections we found that it is typically 
smaller than the LQ signal by two orders of magnitude.
Related to this process is the direct production of a 
quark pair via two photon fusion
\begin{equation}
e + \gamma \to e + q  + \bar{q}.
\end{equation}
However, this process is dominated by the collinear divergence which 
is actually well described by the resolved photon process $eq\to eq$ 
given above.  Once this contribution is subtracted away the remainder 
of the cross section is too small to be a concern \cite{DGP}.
Another possible background consists of $\tau$'s pair produced via 
various mechanisms with one $\tau$ decaying leptonically and the other 
decaying hadronically.  Because of the neutrinos in the final state it  
is expected that the electron and jet's $p_T$ do not in general 
balance which would distinguish these backgrounds from the signal.
However,  this background should be checked in a realistic 
detector Monte Carlo to be sure.
The remaining backgrounds originate from 
heavy quark pair production with one quark decaying 
semileptonically and only the lepton being observed with the 
remaining heavy quark not being identified as such. All such 
backgrounds are significantly smaller than our signal in the kinematic 
region we are concerned with.

\section{Leptoquark Discovery Limits}

In Fig.~2 we show the cross sections for a $\sqrt{s}=1$~TeV $e^+e^-$ 
operating in both the backscattered laser $e\gamma$ mode and in the 
$e^+e^-$ mode.  
The cross section for leptoquarks coupling to the $u$ quark is larger 
than those coupling to the $d$ quark.  This is due to the larger $u$ 
quark content of the photon compared to the $d$ quark content which 
can be traced to the larger $Q_q^2$ of the $u$-quark.  
There exist several different quark distribution functions in the 
literature \cite{nic,DO,DG,GRV,LAC}.  
For the four different leptoquark 
charges we show curves for three different 
distributions functions: Drees and Grassie (DG)\cite{DG}, Gl\"uck, 
Reya and Vogt (GRV)\cite{GRV}, and Abramowicz, Charchula and Levy 
(LAC) set 1\cite{LAC}.  
The different distributions give almost identical results 
for the $Q_{LQ}=-1/3, \; -5/3$ leptoquarks  
and for the $Q_{LQ}=-2/3, \; -4/3$ leptoquarks 
give LQ cross sections that vary by most a factor 
of two, depending on the kinematic region.
In the remainder of our results we will use 
the GRV distribution functions \cite{GRV} which we take to be 
representative of the quark distributions in the photon.

\begin{figure}[b]
\leavevmode
\centerline{
\epsfig{file=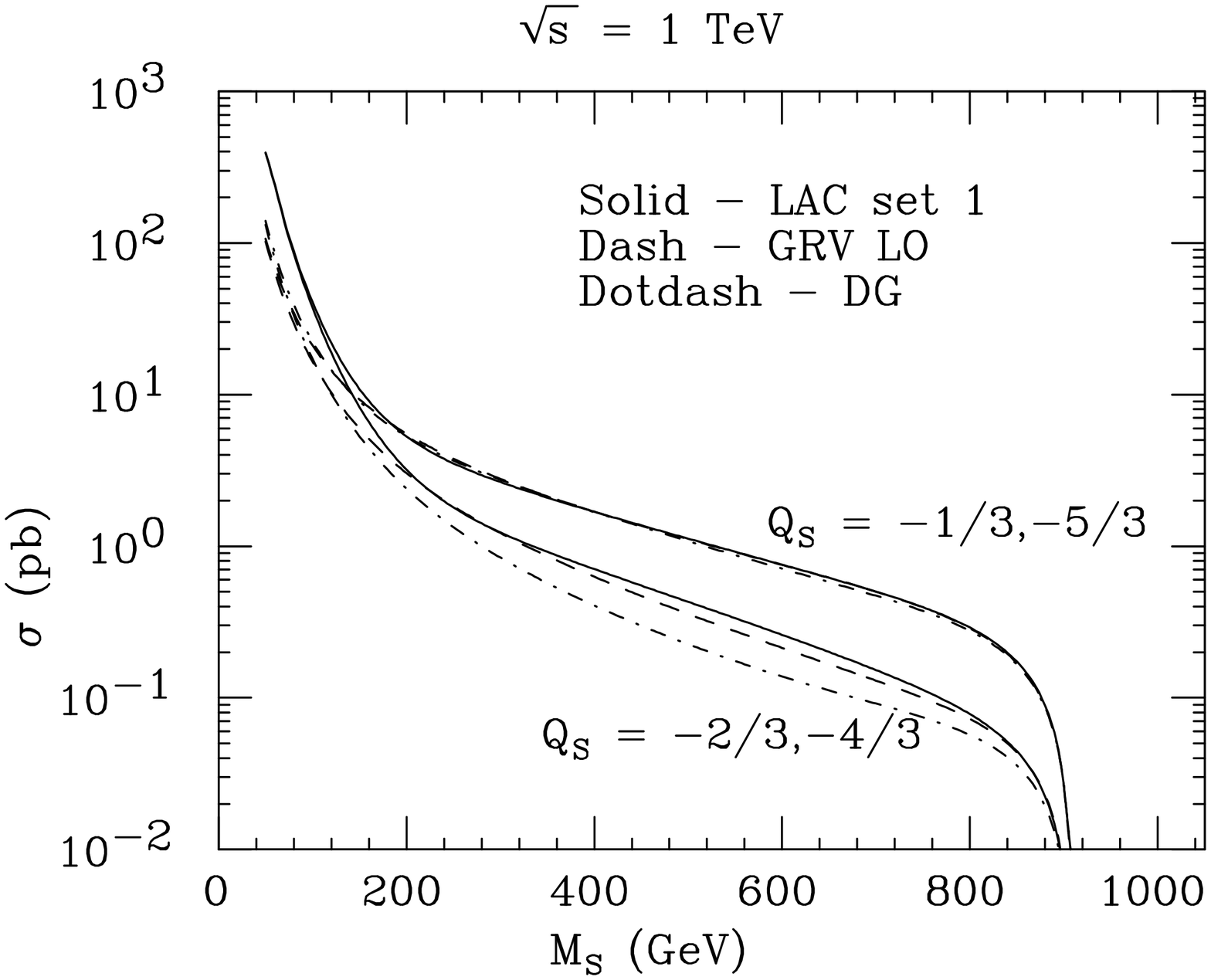,width=8.0cm,clip=}
}
\centerline{
\epsfig{file=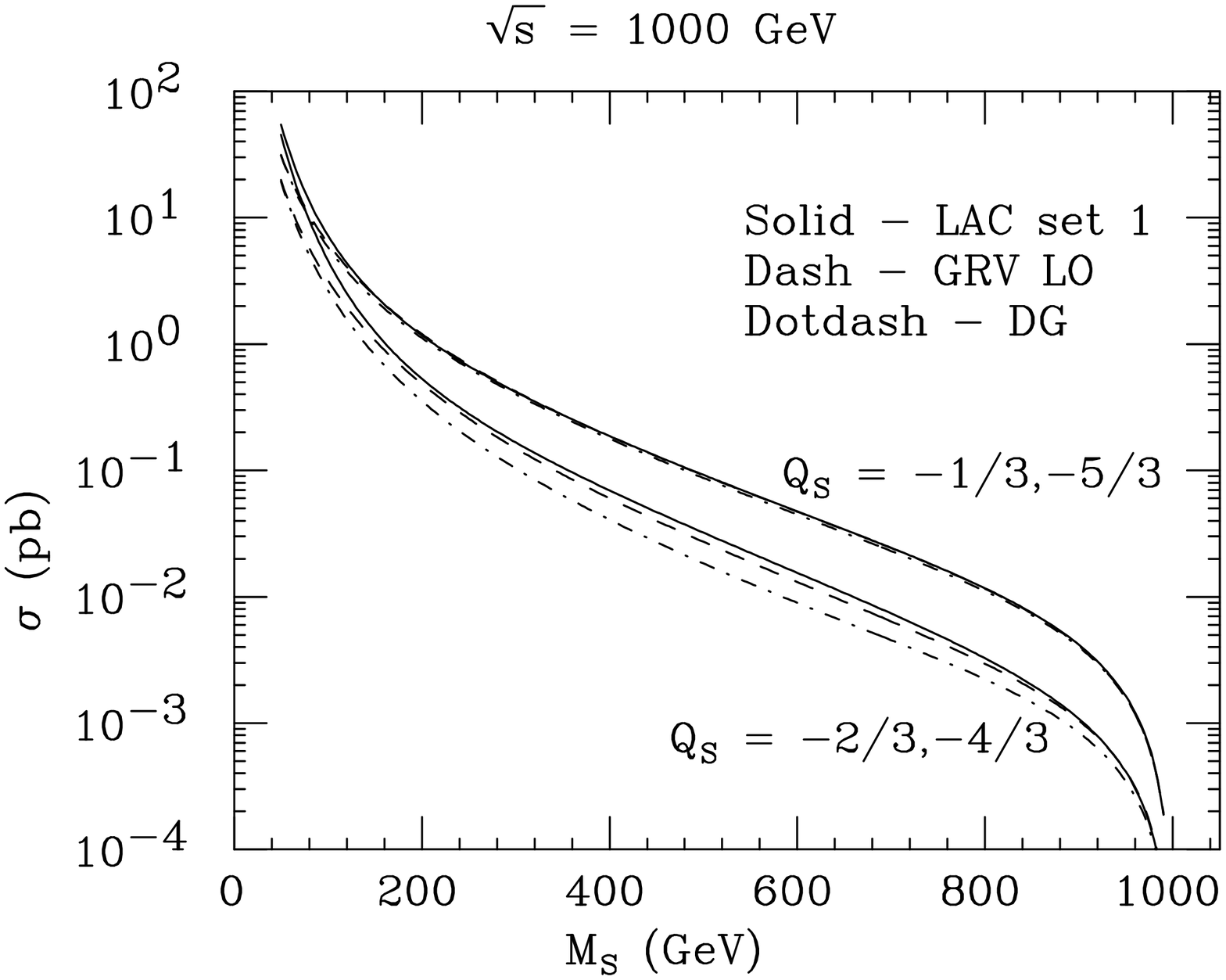,width=8.0cm,clip=}
}
\caption{The cross sections for leptoquark production due to resolved 
photon contributions in $e\gamma$ collisions, with $\kappa$ chosen to 
be 1.  
In the top figure the photon beam is due to laser backscattering in a 
$\sqrt{s}=1000$~GeV $e^+ e^-$ collider.  In the bottom figure the 
the photon distribution is given by the Weizs\"acker-Williams effective 
photon distribution in a $\sqrt{s}=1000$~GeV $e^+ e^-$ collider.
In both cases the 
solid, dashed, dot-dashed line is for resolved photon distribution 
functions of Abramowicz, Charchula and Levy \cite{LAC},  
Gl\"uck, Reya and Vogt \cite{GRV}, Drees and Grassie \cite{DG}, 
respectively.}
\end{figure}

Comparing the cross sections for the two collider modes we see that 
the kinematic limit for the $e\gamma$ mode is slightly lower than the 
$e^+e^-$ mode.  This is because the backscattered laser mode has an 
inherent energy limit beyond which the laser photons pair produce 
electrons.  On the other hand the backscattered laser cross sections 
is larger than the $e^+e^-$ mode.  This simply reflects that the 
backscattered laser photon spectrum is harder than the 
Weizacker-Williams photon spectrum.  To determine the leptoquark 
discovery limits for a given collider we multiply the cross section by 
the integrated luminosity and use the criteria that a certain number 
of signature events would constitute a leptoquark discovery.  When we 
do this we find that the discovery limits for the $e\gamma$ and 
$e^+e^-$ modes are not very different.  Although the $e\gamma$ mode 
has a harder photon spectrum, the $e^+e^-$ mode has a higher kinematic 
limit.  

In Fig. 3 we plot the number of events for various collider energies 
for $e^+e^-$, $e\gamma$, and $\mu^+\mu^-$ colliders.  For the 
$\mu^+\mu^-$ collider we used the $c$ and $s$ quark distributions in 
the photon rather than the $u$ and $d$ quark distributions since we 
only consider generation diagonal leptoquark couplings.  For 
$\sqrt{s}=500$~GeV a $e^+e^-$ collider will have about a 25\% 
higher reach than a $\mu^+\mu^-$ collider due to the larger $u$ and 
$d$ distributions arising from the smaller quark masses.  For the 
highest energy lepton colliders considered the differences become 
relatively small.  For the high luminosities being envisaged, the 
limiting factor in producing enough leptoquarks to meet our discovery 
criteria is the kinematic limit.  Because, for a given $e^+e^-$ centre 
of mass energy, an $e^+e^-$ collider will have a higher energy than an 
$e\gamma$ collider using a backscattered laser, the $e^+e^-$ collider 
will have a higher discovery limit.  Finally, note that the discovery 
limit for  vector leptoquarks is slightly higher than the discovery 
limit for scalar leptoquarks.  This simply reflects the fact that the 
cross section for vector leptoquarks is a factor of two larger than 
the cross section for scalar leptoquarks.  We summarize the discovery 
limits for the various colliders in Table 1.

\begin{figure}[b]
\leavevmode
\centerline{
\begin{turn}{90}
\epsfig{file=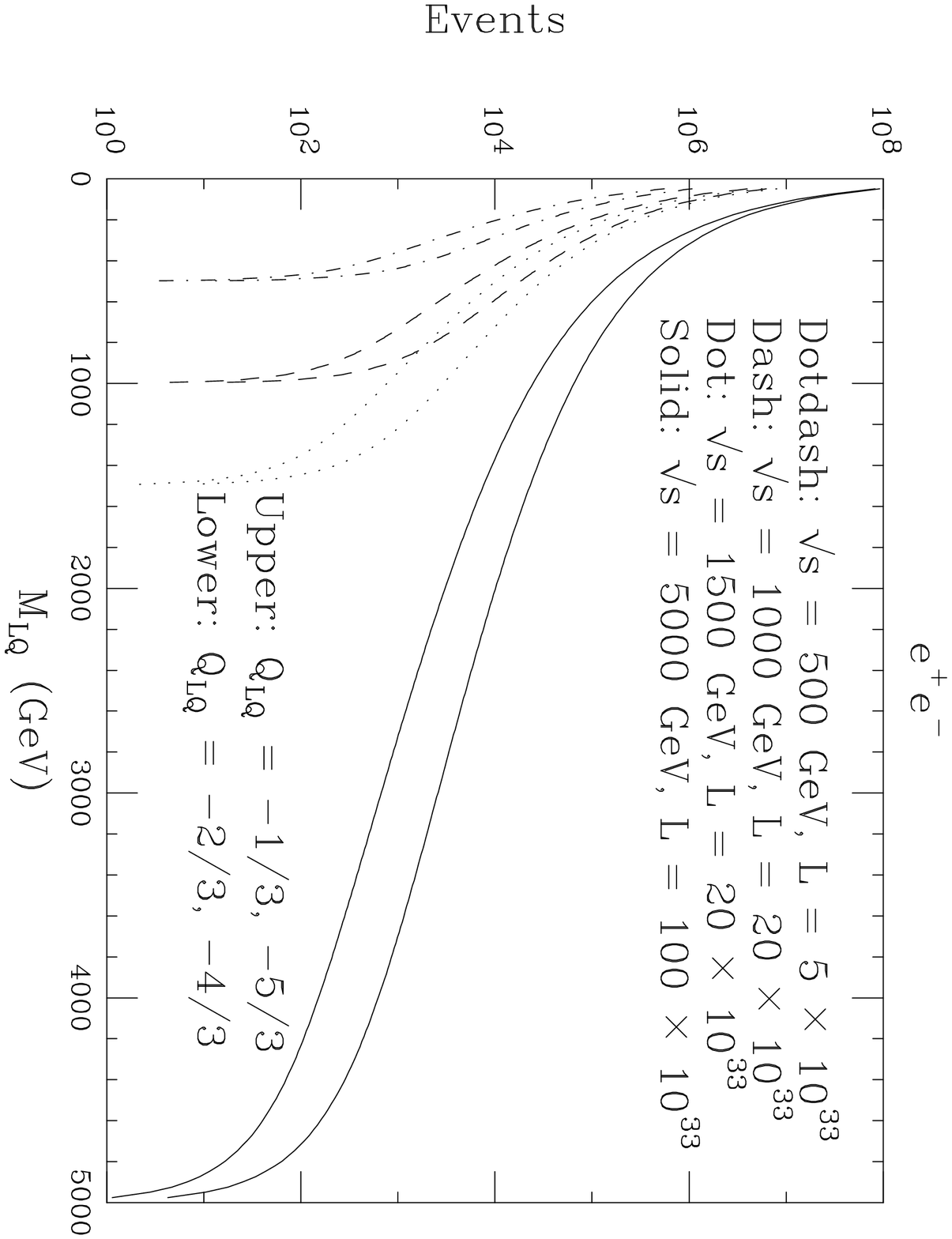,width=5.5cm,clip=}
\end{turn}
}
\vskip 0.2cm
\centerline{
\begin{turn}{90}
\epsfig{file=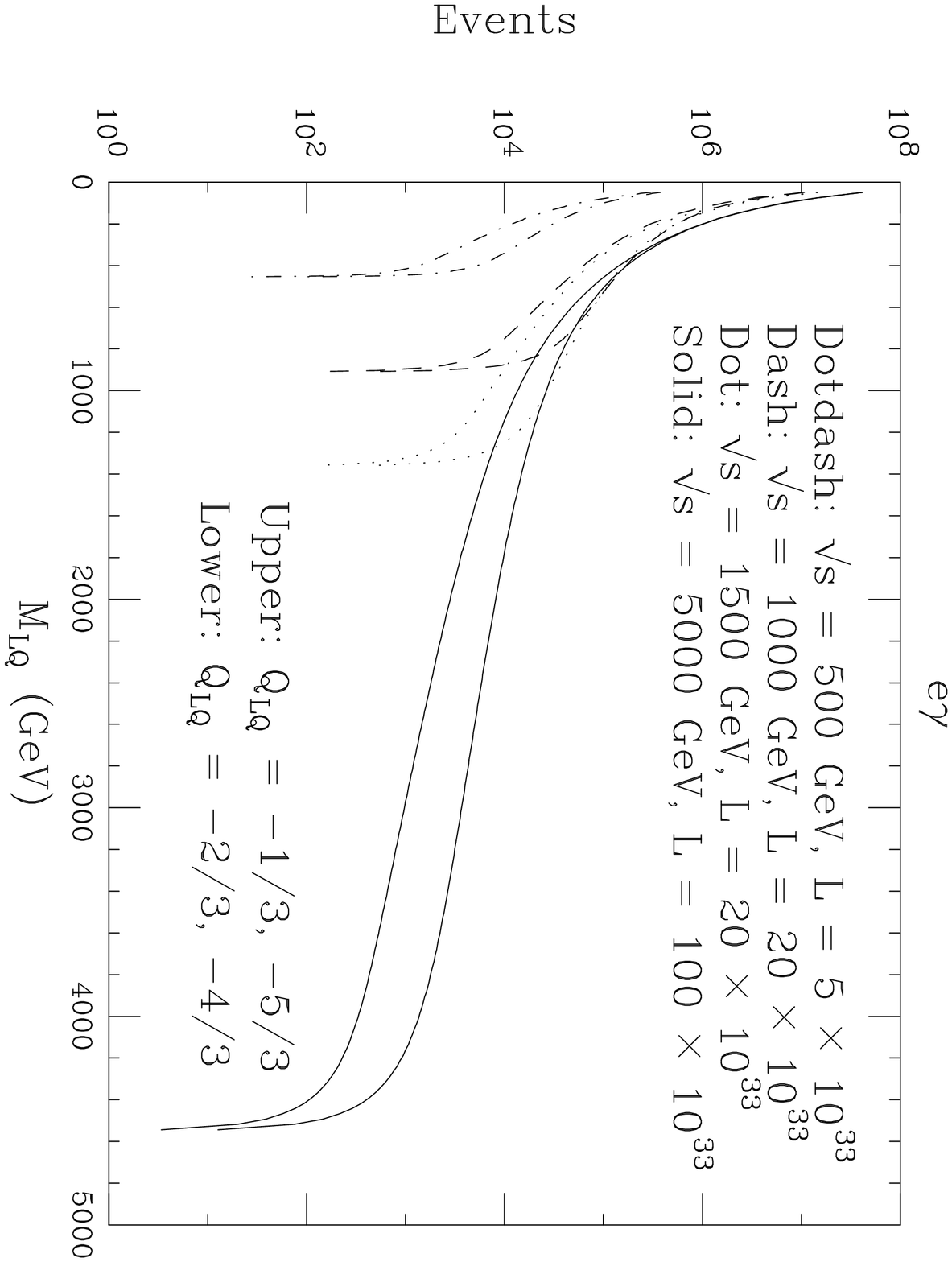,width=5.5cm,clip=}
\end{turn}
}
\vskip 0.2cm
\centerline{
\begin{turn}{90}
\epsfig{file=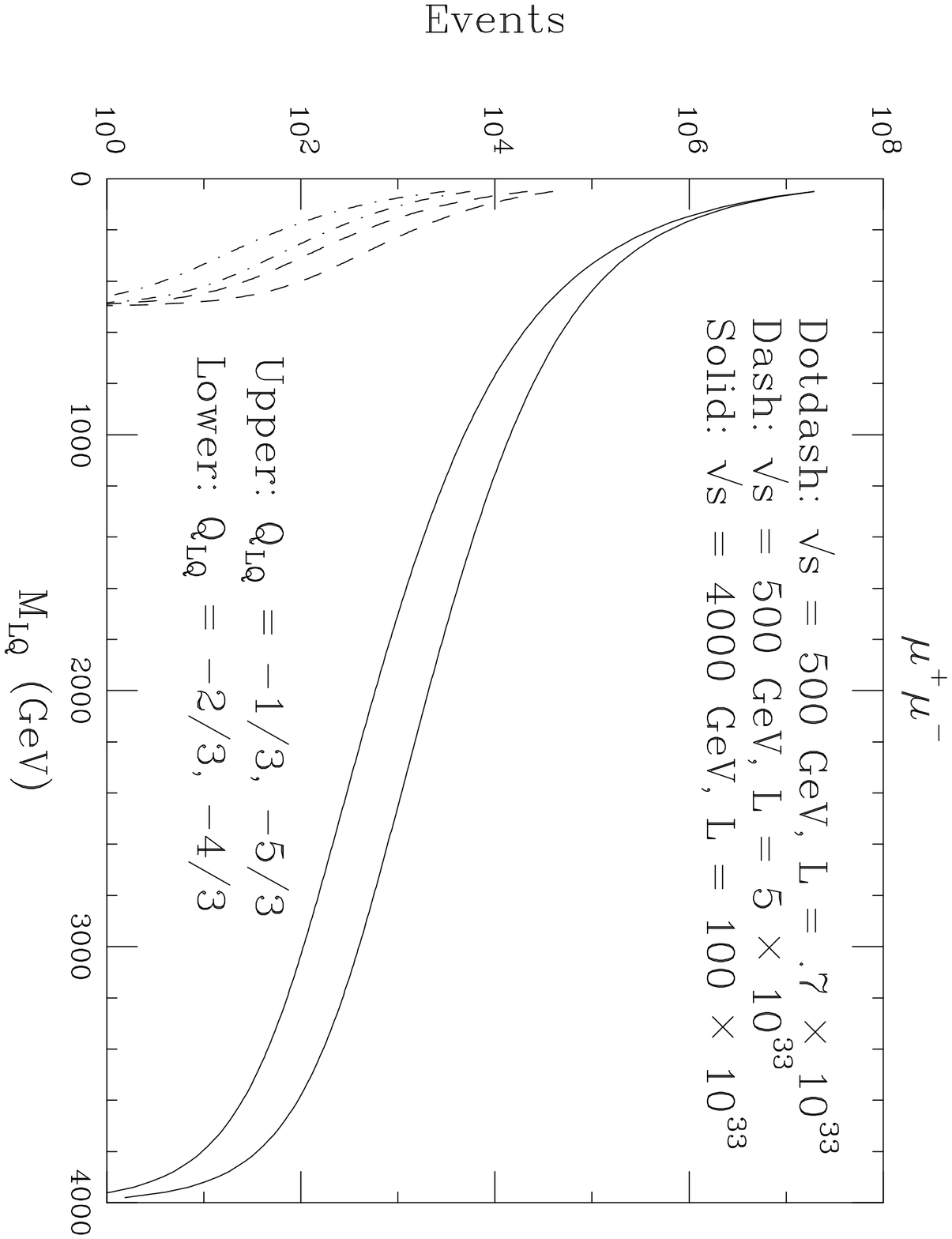,width=5.5cm,clip=}
\end{turn}
}
\caption{Event rates for single leptoquark production in
$e^+e^-$, $e\gamma$, and $\mu^+\mu^-$ collisions.  
The centre of mass energies 
and integrated luminosities are given by the line labelling
in the figures.
The results were obtained using the GRV distribution functions 
\cite{GRV}.}
\end{figure}

\begin{table*}[t]
\begin{center}
\caption{Leptoquark discovery limits for $e^+e^-$, 
$e\gamma$, and $\mu^+\mu^-$ colliders.  The discovery limits are based 
on the production of 100 LQ's for the centre of mass energies 
and integrated luminosities given in columns one and two.
The results were obtained using the GRV distribution functions 
\cite{GRV}.}
\begin{tabular}{llllll}
\hline
\hline
\multicolumn{6}{c}{$e^+e^-$ Colliders}\\ \hline
$\sqrt{s}$ (TeV) & $L$ (fb$^{-1}$) & \multicolumn{2}{c}{Scalar} & 
\multicolumn{2}{c}{Vector}  \\ \hline 
 & & -1/3, -5/3 & -4/3, -2/3 & -1/3, -5/3 & -4/3, -2/3 \\ \hline 
 0.5  & 50 & 490 & 470 & 490 & 480 \\
 1.0  & 200 & 980 & 940 & 980 & 970 \\
 1.5 & 200 & 1440 & 1340 & 1470 & 1410  \\
 5.0 & 1000 & 4700 & 4200 & 4800 & 4500  \\ \hline \hline
\end{tabular}
\vskip 0.6cm
\begin{tabular}{llllll}
\hline
\hline
\multicolumn{6}{c}{$e\gamma$ Colliders} \\ \hline
$\sqrt{s}$ (TeV) & $L$ (fb$^{-1}$) & \multicolumn{2}{c}{Scalar} & 
\multicolumn{2}{c}{Vector}  \\ \hline 
 & & -1/3, -5/3 & -4/3, -2/3 & -1/3, -5/3 & -4/3, -2/3 \\ \hline 
 0.5  & 50 & 450 & 450 & 450 & 440 \\
 1.0  & 200 & 900 & 900 & 910 & 910 \\
 1.5 & 200 & 1360 & 1360 & 1360 & 1360  \\
 5.0 & 1000 & 4500 & 4400 & 4500 & 4500  \\ \hline \hline
\end{tabular}
\vskip 0.6cm
\begin{tabular}{llllll}
\hline
\hline
\multicolumn{6}{c}{$\mu^+\mu^-$ Colliders} \\ \hline
$\sqrt{s}$ (TeV) & $L$ (fb$^{-1}$) & \multicolumn{2}{c}{Scalar} & 
\multicolumn{2}{c}{Vector} \\ \hline 
 & & -1/3, -5/3 & -4/3, -2/3 & -1/3, -5/3 & -4/3, -2/3 \\ \hline 
 0.5  & 0.7 & 250 & 170 & 310 & 220 \\
 0.5  & 50 & 400 & 310 & 440 & 360 \\
 4.0 & 1000 & 3600 & 3000 & 3700 & 3400  \\ \hline
\hline 
\end{tabular}
\end{center}
\end{table*}

\section{Leptoquark Identification}

If a leptoquark were actually discovered the next step would be to 
determine its properties so that we could determine which model it 
originated from.  We will assume that a 
peak in the $e + jet$ invariant mass is observed in some 
collider ({\it i.e.}, the existance of a LQ has been established), 
and so we need simply to identify the particular type of LQ. We assume 
that the leptoquark charge has not been determined and assume no 
intergenerational couplings. Furthermore, we will 
assume that only one of the ten possible types of LQs is present.  
Table~2 of BRW \cite{buch}
gives information on the couplings to various quark and lepton 
combinations; the missing (and necessary) bit of information in BRW
is that the quark 
and lepton have the same helicity (RR or LL) for scalar LQ production while 
they have opposite helicity (RL or LR) for vector LQ production.  It is then 
possible to construct the cross sections for the various helicity combinations 
and consequently the double spin asymmetry \cite{DG2}, 
for the different types of LQs.

Thus, a first step in identifying leptoquarks would be to determine 
the coupling chirality, ie. whether it couples to $e_L$, $e_R$, or 
$e_U$.  This could be accomplished by using electron polarization either 
directly or by using a left-right asymmetry measurement:
\begin{displaymath}
A^{+-} = {{\sigma^+ - \sigma^-}\over{\sigma^+ + \sigma^-}}
= {{C_L^2 -C_R^2}\over {C_L^2 + C_R^2}}
\end{displaymath}
This divides the 10 BRW leptoquark classifications into three groups:
\begin{description}
\item[$e^-_L$:] $\tilde{R}_2$, $S_3$, $U_3$, $\tilde{V}_2$
\item[$e^-_R$:] $\tilde{S}_1$, $\tilde{U}_1$
\item[$e^-_U$:] $U_1$, $V_2$, $R_2$, $S_1$
\end{description}

We can further distinguish whether the leptoquarks are scalar or 
vector. This could be accomplished in two ways.  In the first one can 
study the angular distributions of the leptoquark decay products.  In the 
second we can use the double asymmetry: 
\begin{displaymath}
A_{LL} = {{ (\sigma^{++} + \sigma^{--} ) - (\sigma^{+-} + 
\sigma^{-+} )} \over
{ (\sigma^{++} + \sigma^{--} ) + (\sigma^{+-} + \sigma^{-+} )}}
\end{displaymath}
where the first index refers to the electron helicity and the second 
to the quark helicity.  Because scalars only have a non-zero cross 
section for $\sigma^{++}$ and $\sigma^{--}$ for scalar LQ's
the parton level asymmetry for $e q$ collisions is $\hat{a}_{LL}= +1$.  
Similarly, since vectors only have a 
non-zero cross section for $\sigma^{+-}$ and $\sigma^{-+}$ for vector 
LQ's $\hat{a}_{LL}=-1$.  

To obtain observable
asymmetries one must convolute the parton level cross sections with 
polarized distribution functions.  Doing so will reduce the asymmetries 
from their parton level values of $\pm 1$ so one 
must determine whether the observable asymmetries can distinguish 
between the leptoquark types.  The expressions for the double 
longitudinal spin asymmetry $A_{LL}$ are given in Ref. \cite{DG2}.
In Figures 4 and 5 we plot $A_{LL}$ for 
the $e\gamma$ collider which started with $\sqrt{s}_{e^+e^-}$=1~TeV 
and with
$\sqrt{s}_{e^+e^-}$=5~TeV respectively.  
To obtain these curves we used 
parameterizations of the {\it asymptotic} polarized 
photon distribution functions \cite{hassan,xu}, where it is assumed that 
$Q^2$ and $x$ are large enough that the Vector Meson Dominance part of the 
photon structure is not important, but rather the behavior is dominated by 
the point-like $\gamma q \bar{q}$ coupling. In order to be consistent, we 
used
a similar asymptotic parameterization for the unpolarized photon distribution 
functions as well \cite{nic}, even though various sets of more correct photon 
distribution functions exist ({\it e.g.}, \cite{DO,DG,GRV,LAC}).  We 
only used this asymptotic approximation in the unpolarized case 
for the calculation of the asymmetry, where it is hoped that in taking a 
ratio of the asymptotic polarized to the asymptotic unpolarized photon 
distribution functions, the error introduced will be minimized. Still, we 
suggest that our results be considered cautiously, at least in the relatively 
small LQ mass region.  We note that in the asymptotic 
approximation, the unpolarized photon distribution functions have (not 
unexpectedly) a similar form to the polarized photon distribution 
functions.

In Fig. 4 and 5 we 
show asymmetries for 100\% polarization and for 90\% polarization 
which is considered to be achievable given the SLC experience.  
The error bars are based on an total integrated luminosity of 
200~$fb^{-1}$ for the 1~TeV case and 1000~$fb^{-1}$ for the 5~TeV case.
In 
these figures note that we are showing $-A_{LL}$ for the vector cases 
so that we can use a larger scale.  Quite clearly, polarization would 
enable us to distinguish between vector and scalar.  For the cases 
where there are two types of leptoquarks of the same chiral couplings, 
for example the  scalar isodoublet $\tilde{R}_2$ and scalar isotriplet 
$S_3$ or the vector isoscalar $U_1$ and vector isodoublet $V_2$, we 
could distinguish between them up to about 3/4 the kinematic limit.

\begin{figure*}[b]
\leavevmode
\begin{minipage}[t]{6.0cm}
\centerline{
\begin{turn}{90}
\epsfig{file=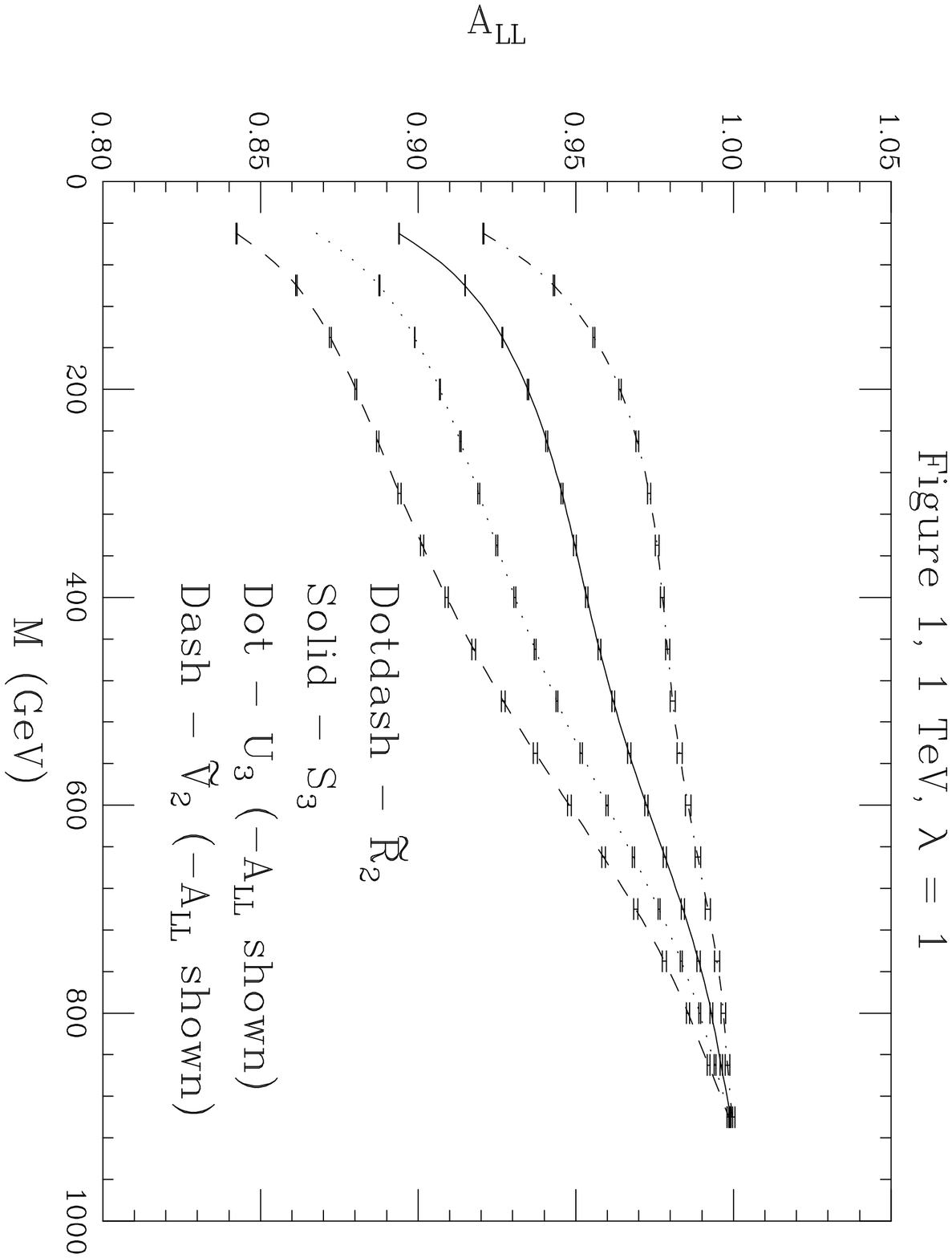,width=4.5cm,clip=}
\end{turn}}
\end{minipage} \
\begin{minipage}[t]{6.0cm}
\centerline{
\begin{turn}{90}
\epsfig{file=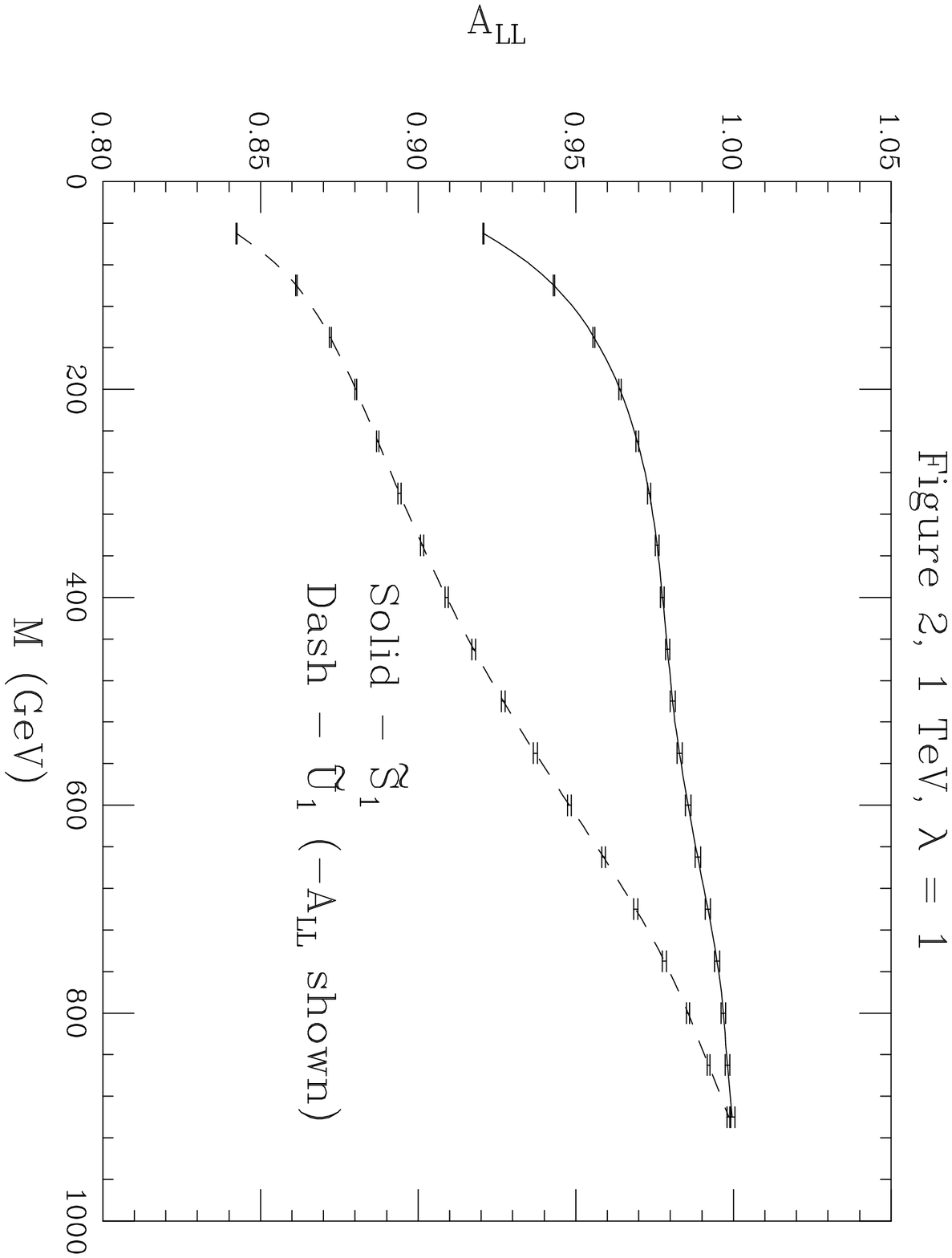,width=4.5cm,clip=}
\end{turn}}
\end{minipage} \
\begin{minipage}[t]{6.0cm}
\centerline{
\begin{turn}{90}
\epsfig{file=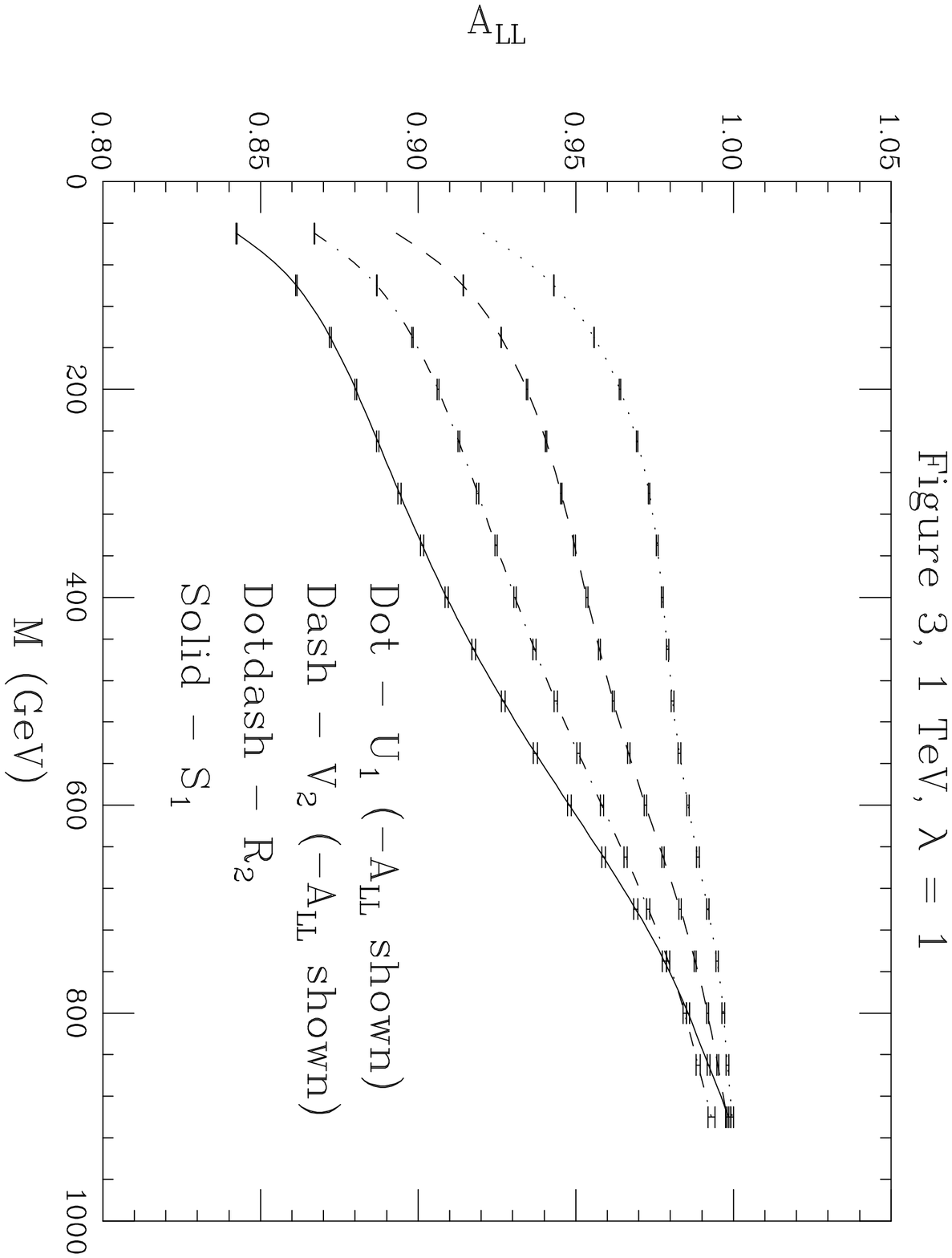,width=4.5cm,clip=}
\end{turn}}
\end{minipage} \\
\begin{minipage}[t]{6.0cm}
\centerline{
\begin{turn}{90}
\epsfig{file=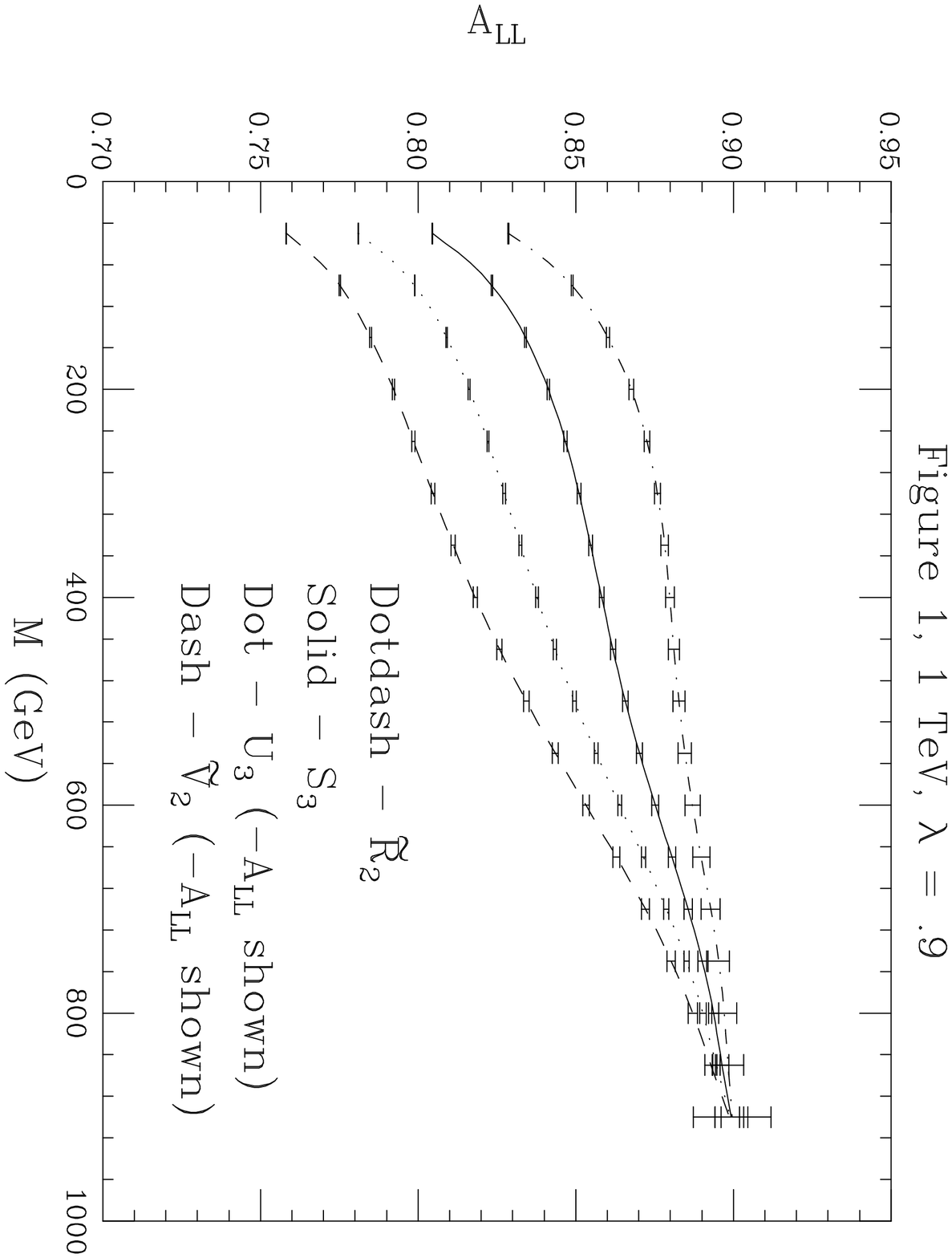,width=4.5cm,clip=}
t\end{turn}}
\end{minipage} \
\begin{minipage}[t]{6.0cm}
\centerline{
\begin{turn}{90}
\epsfig{file=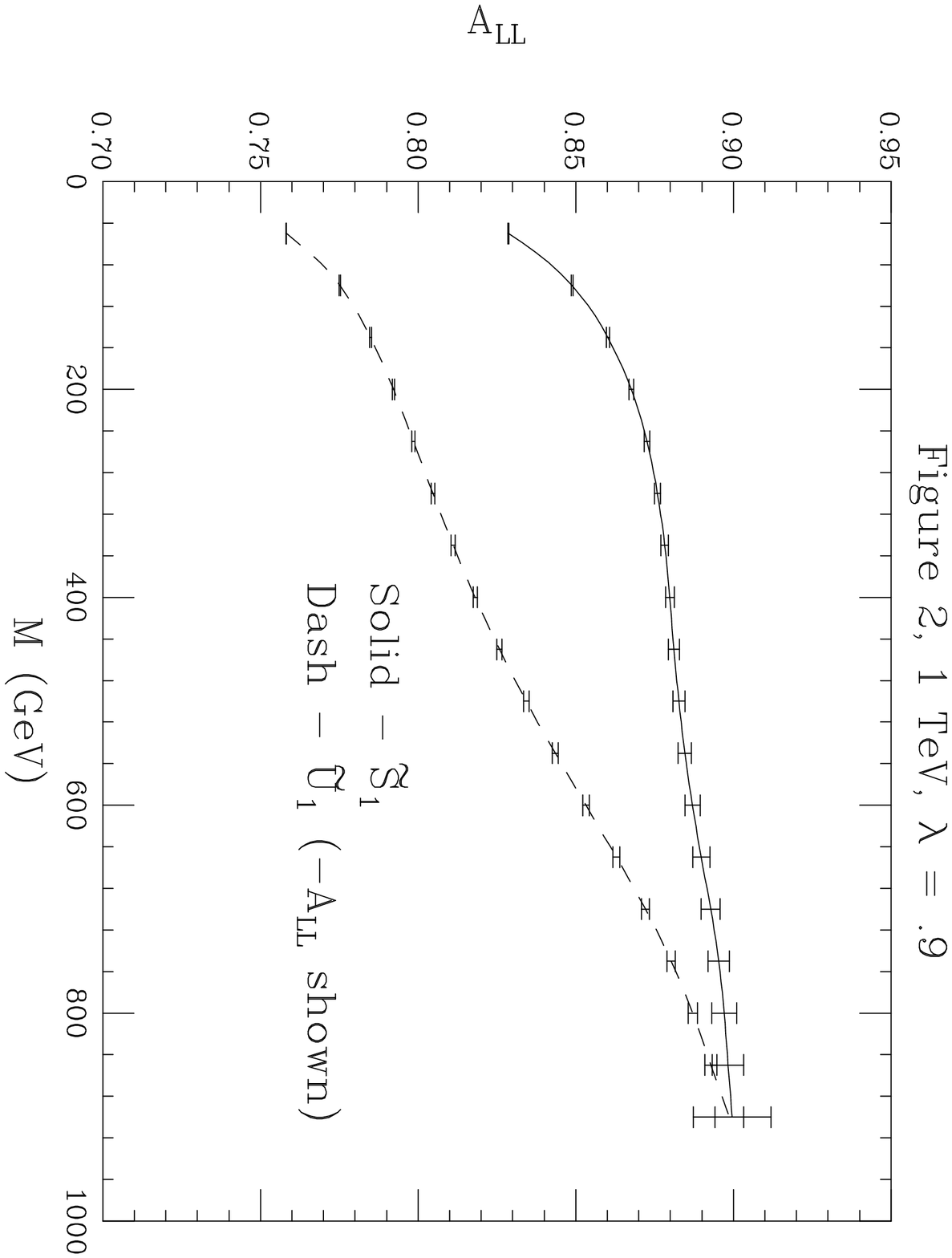,width=4.5cm,clip=}
\end{turn}}
\end{minipage} \
\begin{minipage}[t]{6.0cm}
\centerline{
\begin{turn}{90}
\epsfig{file=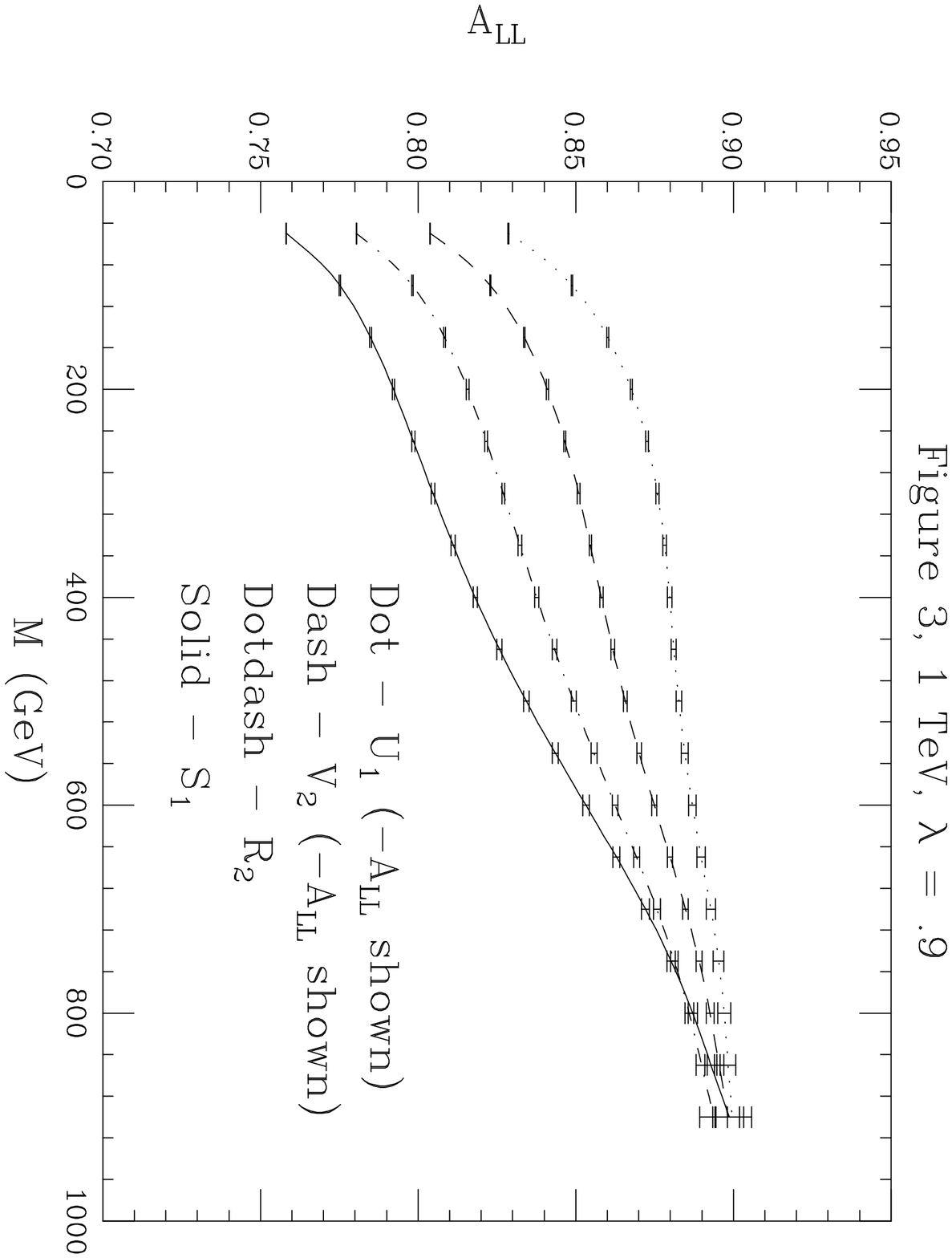,width=4.5cm,clip=}
\end{turn}}
\end{minipage}
\caption{$A_{LL}$ vs $M_{LQ}$ for a 1~TeV $e\gamma$ collider. The 
statistical errors are based on 200~fb$^{-1}$.  The top 
row is for 100\% polarization and the bottom row for 90\% polarization.
The first column is for LQ's which couple only to left-handed 
electrons, the second column is for LQ's which couple only to 
right-handed electrons, and the third column is for LQ's which couple 
to both left and right-handed electrons.}
\end{figure*}

\begin{figure*}[b]
\leavevmode
\begin{minipage}[t]{6.0cm}
\centerline{
\begin{turn}{90}
\epsfig{file=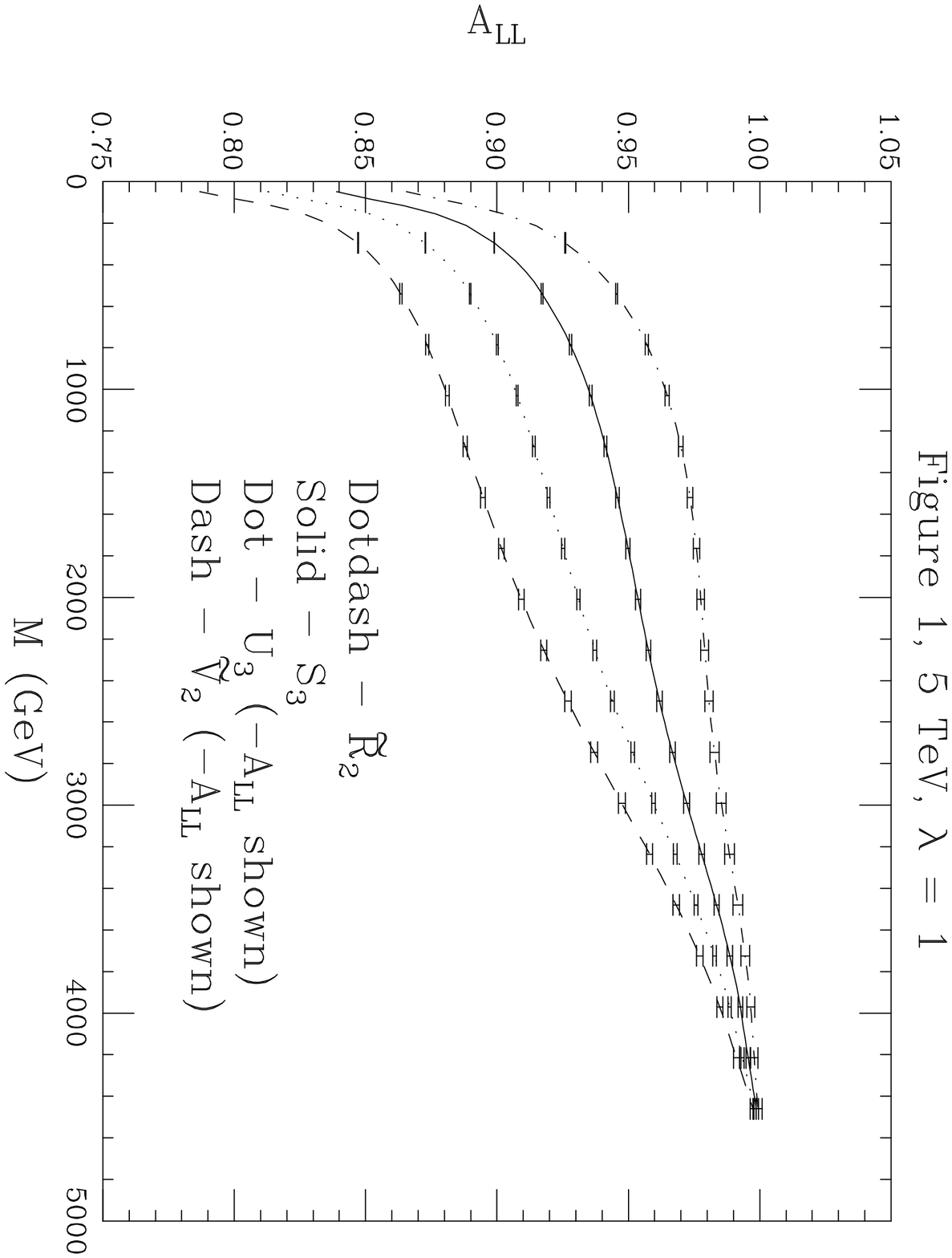,width=4.5cm,clip=}
\end{turn}}
\end{minipage} \
\begin{minipage}[t]{6.0cm}
\centerline{
\begin{turn}{90}
\epsfig{file=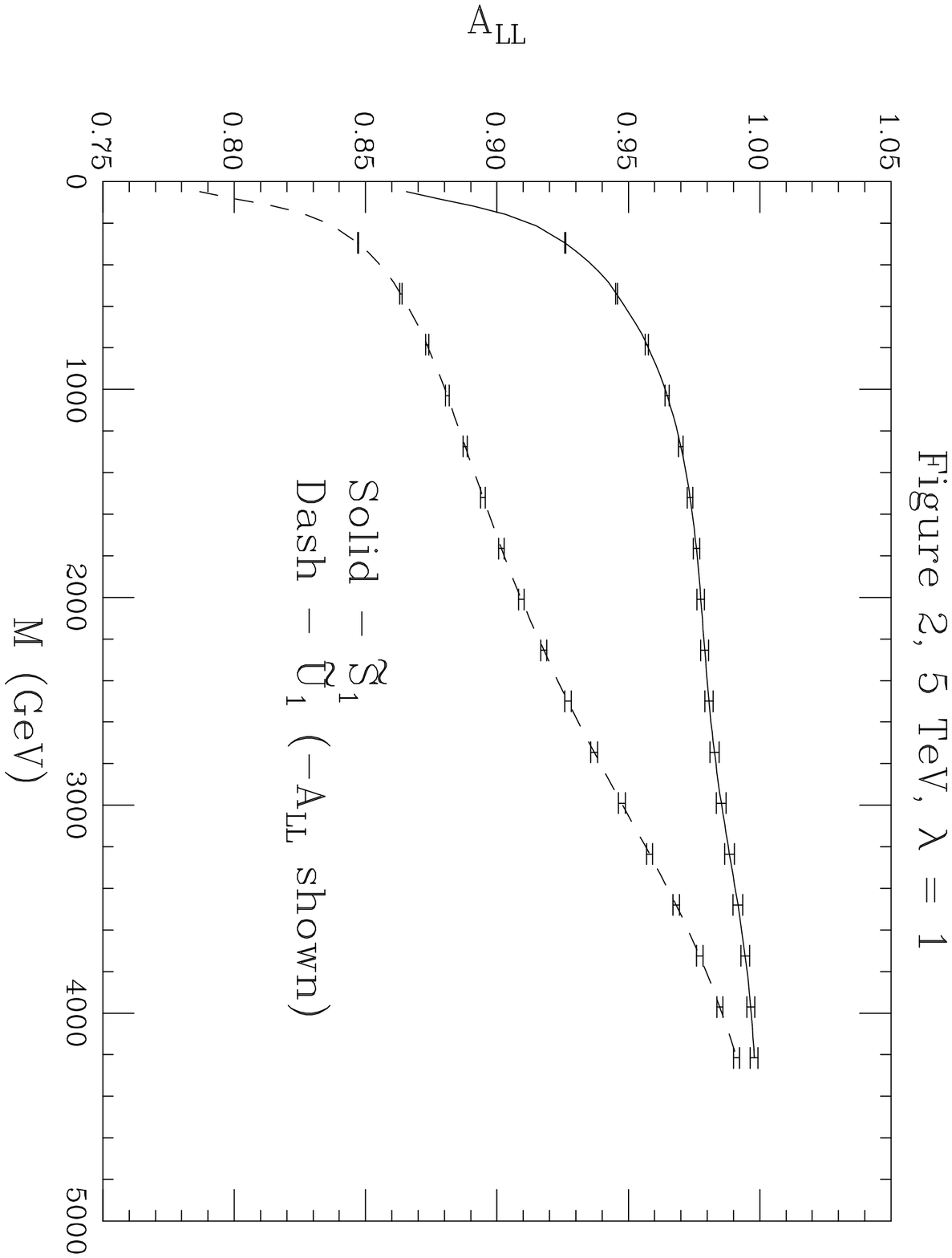,width=4.5cm,clip=}
\end{turn}}
\end{minipage} \
\begin{minipage}[t]{6.0cm}
\centerline{
\begin{turn}{90}
\epsfig{file=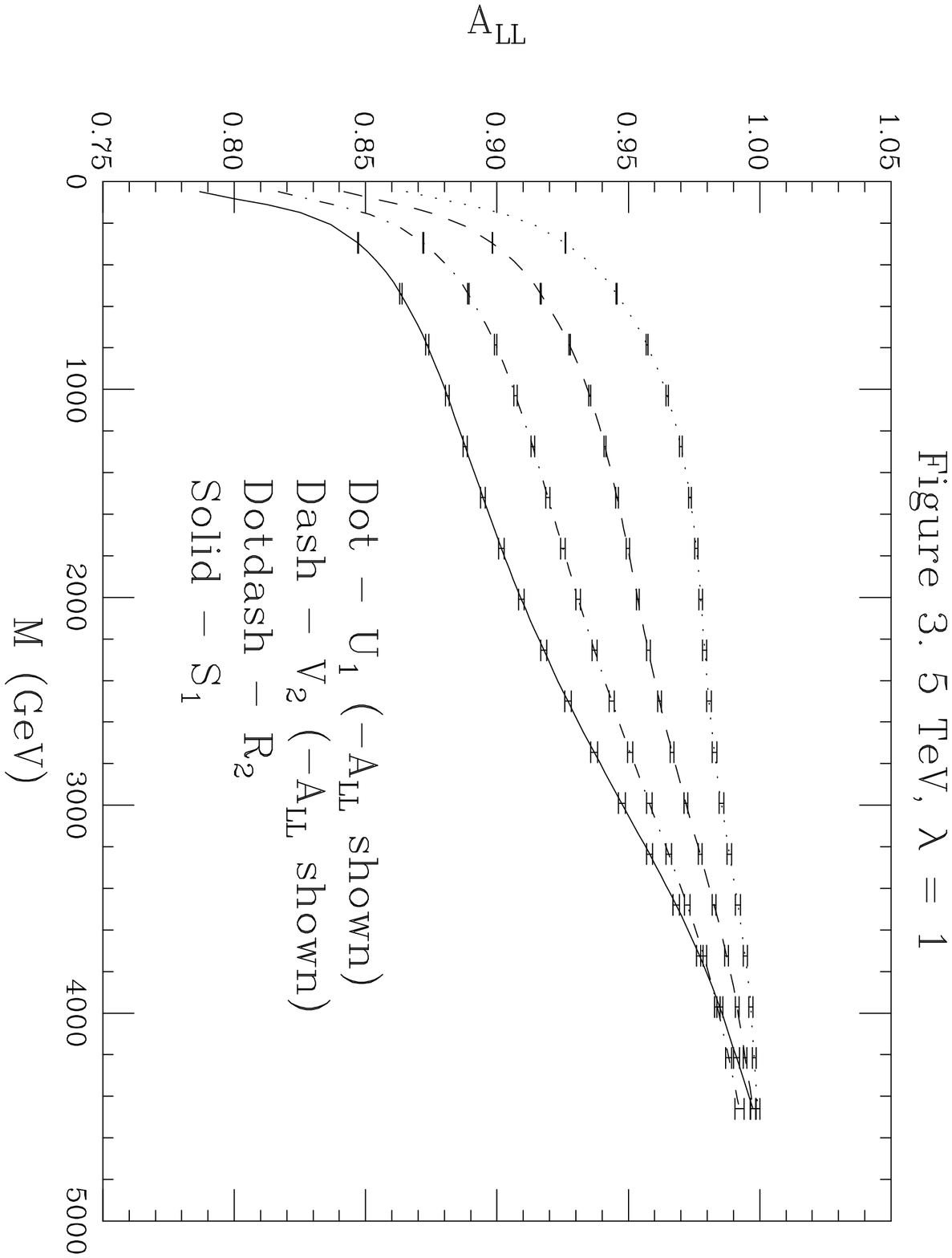,width=4.5cm,clip=}
\end{turn}}
\end{minipage} \\
\begin{minipage}[t]{6.0cm}
\centerline{
\begin{turn}{90}
\epsfig{file=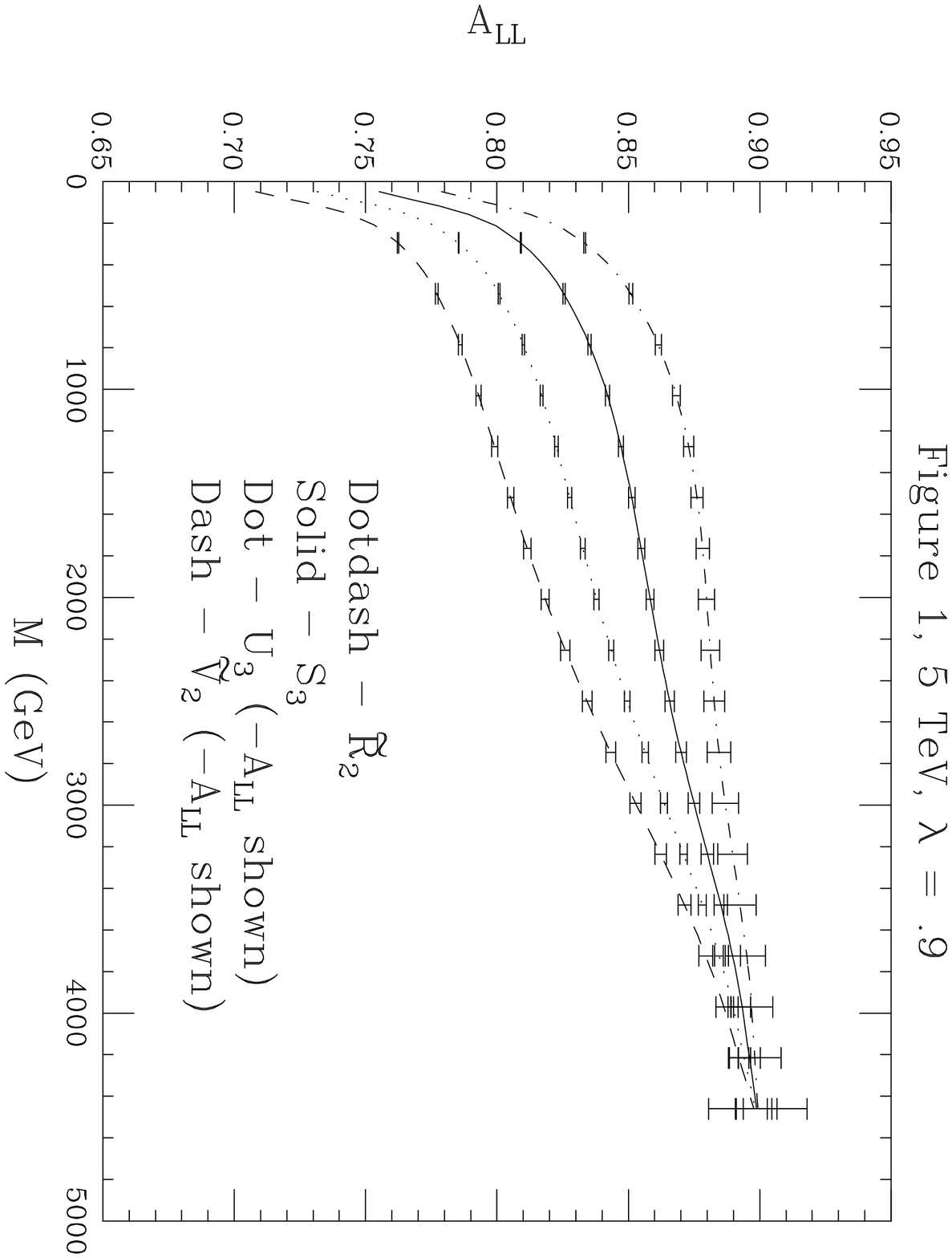,width=4.5cm,clip=}
\end{turn}}
\end{minipage} \
\begin{minipage}[t]{6.0cm}
\centerline{
\begin{turn}{90}
\epsfig{file=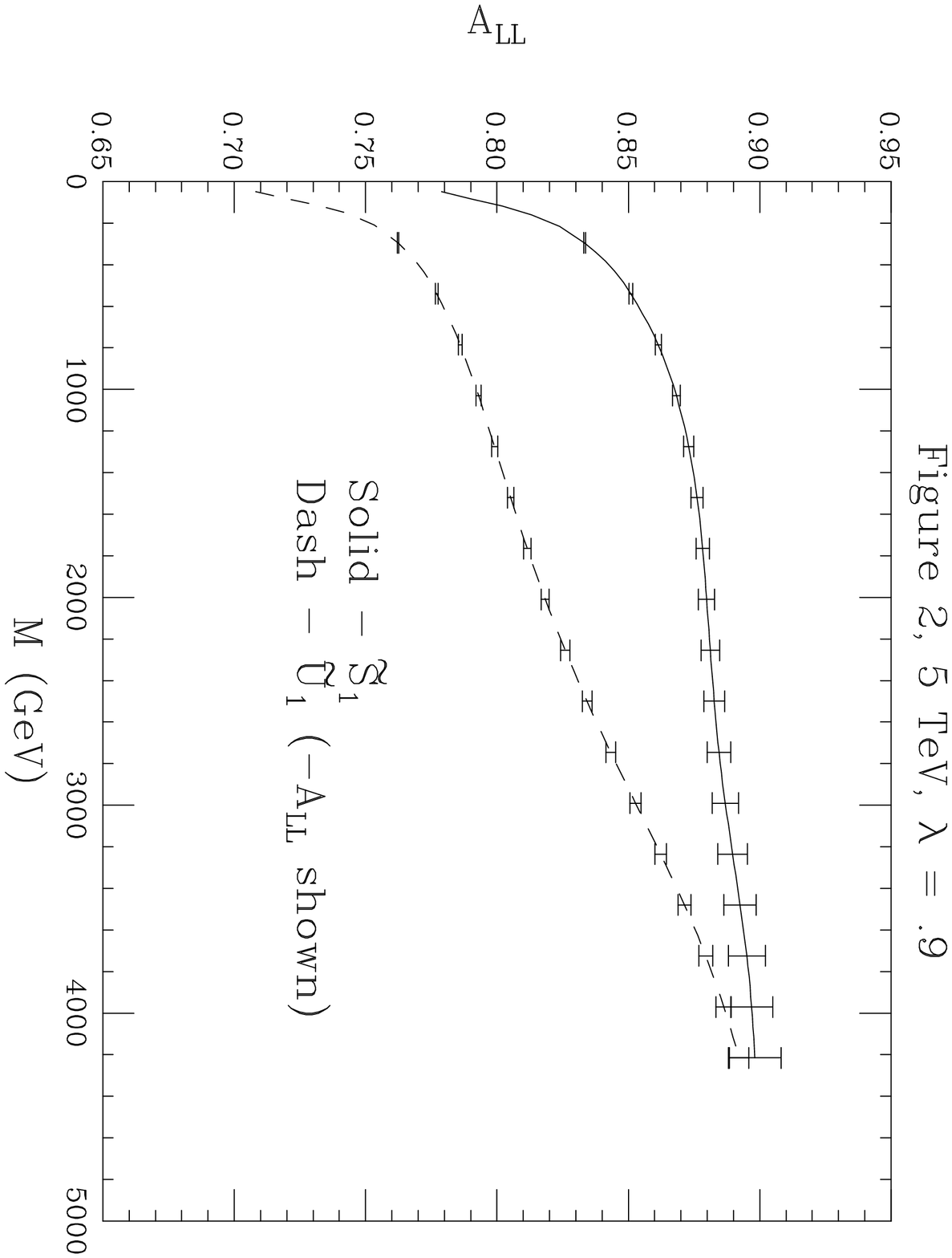,width=4.5cm,clip=}
\end{turn}}
\end{minipage} \
\begin{minipage}[t]{6.0cm}
\centerline{
\begin{turn}{90}
\epsfig{file=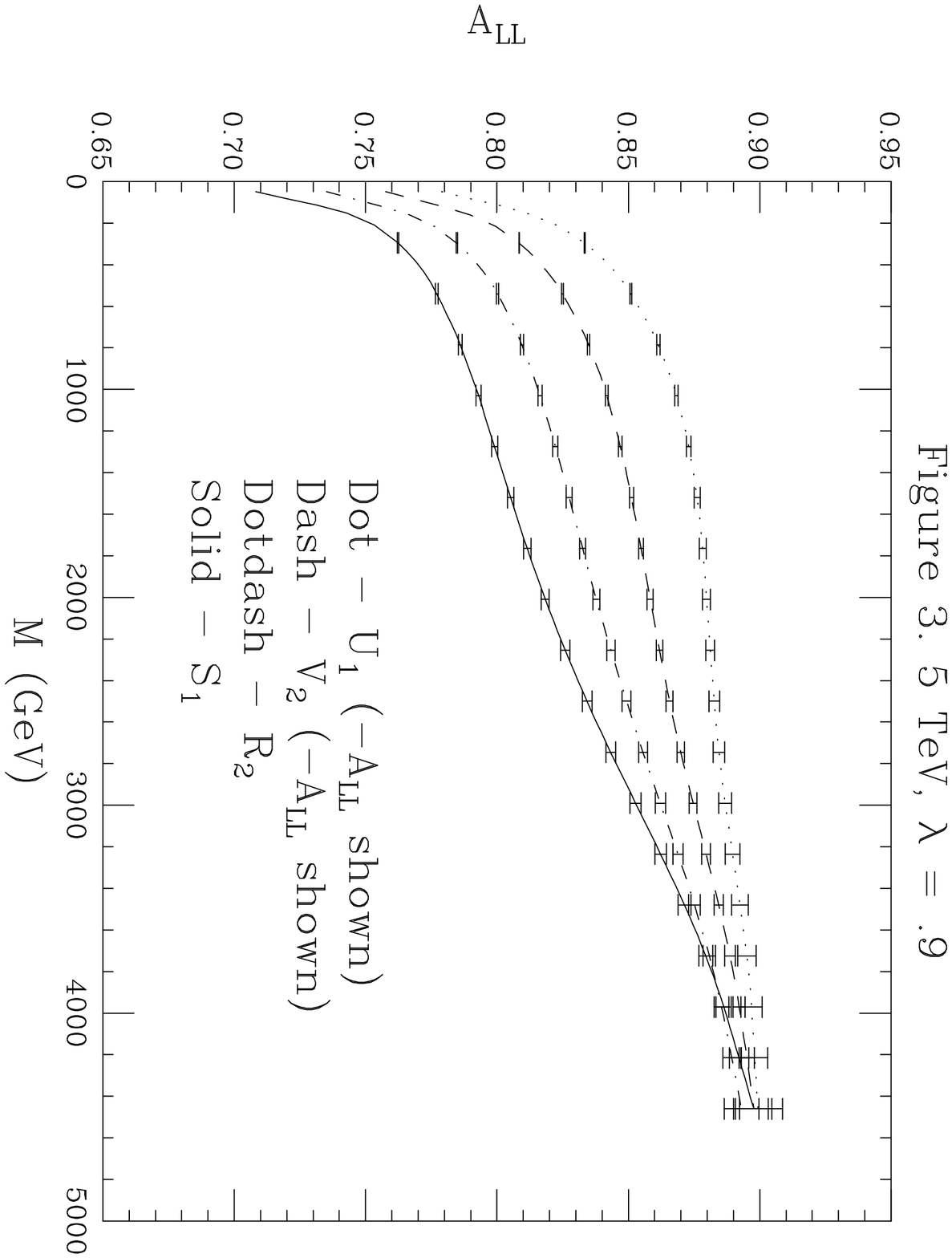,width=4.5cm,clip=}
\end{turn}}
\end{minipage} 
\caption{$A_{LL}$ vs $M_{LQ}$ for a 5~TeV $e\gamma$ collider.  The 
statistical errors are based on 1000~fb$^{-1}$. The top 
row is for 100\% polarization and the bottom row for 90\% polarization.
The first column is for LQ's which couple only to left-handed 
electrons, the second column is for LQ's which couple only to 
right-handed electrons, and the third column is for LQ's which couple 
to both left and right-handed electrons.}
\end{figure*}

Finally, one additional bit of information
to further differentiate among the various possible leptoquarks is to
search for the $\nu q'$ decay mode.  This signature is quite 
similar to a supersymmetric particle decay (high $p_{_T}$ jet plus 
missing $p_{_T}$) so that
although it cannot be used to unambiguously determine 
the existence on leptoquarks, it can be used, in conjunction with the 
observation of an approximately equal number of $e q$ events (as 
expected in some models) to provide further information on leptoquark 
couplings.  Taken together,  the leptoquark type can be 
uniquely determined.  If more than one leptoquark were 
discovered, determining their properties would tell us their 
origin and therefore, the underlying theory.

\section{Summary}

To summarize, we have presented results for single leptoquark 
production in $e\gamma$, $e^+e^-$, and $\mu^+\mu^-$ collisions.
The discovery limits for leptoquarks is very close to the centre of 
mass energy of the colliding particles. It also 
appears that a polarized $e \gamma$ collider can 
be used to differentiate between the different models of LQs that can exist, 
essentially up to the kinematic limit.  Furthermore, it is quite 
easy to distinguish scalar LQs from vector LQs for all LQ mass (given that the 
LQ is kinematically allowed).  Thus $e^+e^-$, $e\gamma$, and 
$\mu^+\mu^-$ colliders have much to offer in the searches for 
leptoquarks.  If leptoquarks were discovered, $e\gamma$ colliders 
could play a crucial role in unravelling their properties, and 
therefore the underlying physics.

%

\end{document}